# $^{23}$Na NMR spin-lattice relaxation reveals ultrafast Na$^+$ ion dynamics in the solid electrolyte Na$_{3.4}$Sc$_{0.4}$Zr$_{1.6}$(SiO$_4$)$_2$PO$_4$

S. Lunghammer,[1] D. Prutsch,[1] S. Breuer,[1] D. Rettenwander, I. Hanzu,[1,2] Q. Ma,[3] F. Tietz,[3,4] and M. Wilkening[1,2]

[1] Institute for Chemistry and Technology of Materials (NAWI Graz), and Christian Doppler Laboratory for Batteries, Stremayrgasse 9, Graz University of Technology, A-8010 Graz (Austria).

[2] ALISTORE-ERI European Research Institute, 33 Rue Saint Leu, F-80039 Amiens (France).

[3] Forschungszentrum Jülich GmbH, Institute of Energy and Climate Research, Materials Synthesis and Processing (IEK-1), D-52425 Jülich (Germany)

[4] Helmholtz-Institute Münster, c/o Forschungszentrum Jülich GmbH, D-52425 Jülich (Germany).

## Abstract

The realization of green and economically friendly energy storage systems needs materials with outstanding properties. Future batteries based on Na as an abundant element take advantage of non-flammable ceramic electrolytes with very high conductivities. Na$_3$Zr$_2$(SiO$_4$)$_2$PO$_4$-type superionic conductors are expected to pave the way for inherently safe and sustainable all-solid-state batteries. So far, only little information has been extracted from spectroscopic measurements to clarify the origins of fast ionic hopping on the atomic length scale. Here we combined broadband conductivity spectroscopy and nuclear magnetic resonance (NMR) relaxation to study Na ion dynamics from the μm to the angstrom length scale. Spin-lattice relaxation NMR revealed a very fast Na ion exchange process in Na$_{3.4}$Sc$_{0.4}$Zr$_{1.6}$(SiO$_4$)$_2$PO$_4$ that is characterized by an unprecedentedly high self-diffusion coefficient of $9 \times 10^{-12}$ m$^2$s$^{-1}$ at −10 °C. Thus, well below ambient temperature the Na ions have access to elementary diffusion processes with a mean residence time $\tau_{NMR}$ of only 2 ns. The underlying asymmetric diffusion-induced NMR rate peak and the corresponding conductivity isotherms measured in the MHz range reveal correlated ionic motion. Obviously, local but extremely rapid Na$^+$ jumps, involving especially the transition sites in Sc-NZSP, trigger long-range ion transport and push ionic conductivity up to 2 mS/cm at room temperature.

# $^{23}$Na NMR spin-lattice relaxation reveals ultrafast Na$^+$ ion dynamics in the solid electrolyte Na$_{3.4}$Sc$_{0.4}$Zr$_{1.6}$(SiO$_4$)$_2$PO$_4$

S. Lunghammer,[1] D. Prutsch,[1] S. Breuer,[1] D. Rettenwander[1], I. Hanzu,[1,2] Q. Ma,[3] F. Tietz,[3,4] and M. Wilkening[1,2]

[1] Institute for Chemistry and Technology of Materials (NAWI Graz), and Christian Doppler Laboratory for Batteries, Stremayrgasse 9, Graz University of Technology, A-8010 Graz (Austria).

[2] ALISTORE-ERI European Research Institute, 33 Rue Saint Leu, F-80039 Amiens (France).

[3] Forschungszentrum Jülich GmbH, Institute of Energy and Climate Research, Materials Synthesis and Processing (IEK-1), D-52425 Jülich (Germany)

[4] Helmholtz-Institute Münster, c/o Forschungszentrum Jülich GmbH, D-52425 Jülich (Germany).

In the search of safe and long-lasting energy storage systems all-solid-state batteries entered the spotlight of research [1]. One of the major outstanding challenges is to develop stable electrolytes with extraordinarily high ion conductivities at ambient conditions [2]. Since no flammable liquids are used in such batteries they are regarded as inherently safe and might easily withstand high operation temperatures [3]. In recent years, we witnessed the discovery of a couple of very fast and promising Li$^+$ ion conductors [4-10]. The market availability of Li might, however, narrow the widespread realization of large scale stationary applications. Sustainable batteries [11, 12] with Na$^+$ as ionic charge carrier [3, 13-15] do not carry this risk as Na is abundantly available in the earth's crust. The cost-effective realization of environmentally benign ceramic batteries, however, will need solid electrolytes with a very high ion mobility to compete with the much faster dynamic processes in the liquid state [16, 17].

Although moderate to very fast ion conducting Na compounds have so far been frequently presented [15, 16, 18-21]; ion dynamics in non-sulfidic, *i.e.*, air-insensitive, Na$^+$ electrolytes with extremely high ion conductivities exceeding 1 mS/cm at room temperature have, apart from the well-known Na-$\beta''$-alumina [12], only very rarely been characterized in detail [16, 18, 22]. Here we present an in-depth study to explore the roots of extremely rapid 3D ion dynamics in Na$_{3.4}$Sc$_{0.4}$Zr$_{1.6}$(SiO$_4$)$_2$PO$_4$ (Sc-NZSP), which was prepared via a wet-chemical route [18]. The parent compound Na$_3$Zr$_2$(SiO$_4$)$_2$PO$_4$ (NZSP), following the pioneering work of Hong and Goodenough [23, 24], served as benchmark. It has been shown recently that substitution of Zr by Sc in NZSP leads to an increase of ionic conductivity [22]. For the reason of charge compensation, the replacement of parts of Zr ions with an equimolar amount of Sc ions increases the number density of (mobile) Na ions. Sc-NZSP indeed exhibited Na ion bulk conductivities $\sigma'$ of 2 mS/cm at room temperature with activation energies ranging from 0.13 eV to 0.31 eV. Although in the Sc-free compound $\sigma'$ was somewhat lower (1 mS/cm) the grain-boundaries in NZSP turned out to be much less blocking for the Na$^+$ ions. This feature ensured facile ion transport over distances in the μm range. Such long-range ion transport is above all the most important property for any battery application [17].

It is beyond any doubt that reporting such high conductivities [4-8, 10, 25-28], especially for sodium-based batteries [19-21, 29, 30], is of enormous importance. Understanding their roots, preferably with the help of theory [31, 32], is also crucial and would enable us to safely control their dynamic properties. Yet, for many materials, however, there is still no complete picture available consistently describing the interrelation of the elementary steps of hopping with long-range ion transport. In many cases, it is not even clear which charge carrier and dynamic processes are responsible for the high conductivities reported. Such features cannot be probed by conductivity measurements alone. Instead, one should use complementary techniques that can observe dynamics over wide scales in time and length [33-35]. In this study, we combined time-domain nuclear magnetic resonance (NMR) spectroscopy with broadband conductivity spectroscopy to quantify short-range as well as long-range Na$^+$ ion diffusivities. Using only NMR spin-lattice relaxometry working at atomic scale [36, 37] we managed to get a sneak peek at the very rapid exchange processes the Na ions are performing at temperatures very well below ambient. The data collected provide a first experimental rationale for the very high ionic conductivity observed in NZSP-type compounds.

Since our impedance measurements covered a dynamic range of more than ten decades they allowed us i) to separate the bulk from grain boundary (g.b.) response at temperatures where batteries are usually operated and ii) to directly compare electric relaxation with magnetic spin relaxation which was measured in the kHz and GHz range, respectively. Here, NMR relaxometry confirmed the extraordinarily high conductivity of Na$_{3.4}$Sc$_{0.4}$Zr$_{1.6}$(SiO$_4$)$_2$PO$_4$ and, furthermore, pointed to a distribution of activation energies including values as low as 0.13 eV and 0.16 eV. Hence, the ions are subjected not to a single but to various dynamic processes being stepwise activated with temperature or running in parallel. To our knowledge, this dynamic performance, although Na$^+$ ions are of course larger and therefore perhaps more sluggish than Li$^+$ ions, exceeds those of the best Li oxides currently discussed as ceramic electrolytes [2, 38]. Considering the electrochemical stability of NZSP-type materials [39] this behaviour clearly opens the field for the development of powerful Na ion solid-state batteries.

## Results and discussion

Na$_3$Zr$_2$(SiO$_4$)$_2$(PO$_4$) is reported to crystallize with monoclinic symmetry (space group *C*2/*c*, see the views along different crystallographic axes in Figure **1a** to **1c**. In the monoclinic form of NZSP the Na$^+$ ions occupy three crystallographically inequivalent sites 4*d*, 4*e*, and 8*f*, which are partially occupied. The large fraction of vacant Na sites and the favourable connection of their polyhedra ensure rapid exchange of the Na$^+$ ions as discussed in literature [22, 40]. The preparation and characterisation of the samples studied here (NZSP and Sc-NZSP) has already been described elsewhere [18, 22].

**Electrical conductivity: bulk vs. grain boundary responses.** We used a sintered Sc-NZSP pellet (1.1 mm in thickness and 7 mm in diameter) equipped with sputtered ion-blocking electrodes of gold to ensure electrical contact for our conductivity and impedance measurements, see Figs. **1d** to **1g**. Conductivity isotherms showing the real part $\sigma'$ of the complex conductivity as a function of frequency ν are displayed in Fig. **1f**. If recorded over a large frequency range spanning from the μHz to the GHz regime at least two main contributions were revealed; they are labelled (A) and (B) in Fig. **1f**. While they appear as plateaus in the $\sigma'(\nu)$ plot, they manifest as two independent semicircles in the complex plane plots, *i.e.*, Nyquist illustrations, see Fig. **1e** showing the responses recorded at −100 °C with the apex frequencies indicated. As the capacity of plateau (A) is in the nF range we attribute the semicircle in the low frequency range to the g.b. response. The spike seen at very low frequencies is attributed to ionic polarisation effects in front of the Na$^+$ blocking electrodes. Plateau (B) is assigned to the bulk response as the corresponding capacity $C_{bulk}$ is in the pF range which is typical for electrical relaxation processes taking place in the grain.

Most importantly, at room temperature the bulk semicircle (see Fig. **1g**), which was recorded in the GHz regime, corresponds to a specific conductivity in the order of 2 mS/cm. The g.b. response yields values which are two orders of magnitude lower (see also the difference of the plateaus $\sigma'_{bulk}(\nu)$ and $\sigma'_{g.b.}(\nu)$ in Fig. **1f**). The two processes can also be clearly seen if the imaginary part $Z''$ of the complex impedance and that of the complex modulus ($M''$) are analysed. The corresponding curves $-Z''(\nu)$ and $M''(\nu)$ are shown in Fig. **1d** and reveal two peaks that differ in amplitude according to the capacities associated with the two processes they represent. Here we have $C_{g.b.}/C_{bulk} \approx 100$ leading to the same ratio in peak amplitudes $M''_{max./bulk}/M''_{max./g.b.}$. Since $M''_{(max.)} \propto 1/C$, bulk responses are more prominent in the modulus representation, as expected. Almost the same ratio is probed for the real part of the complex permittivity; values in the order of $10^3$ or $10^4$ are typically seen for g.b. contributions.

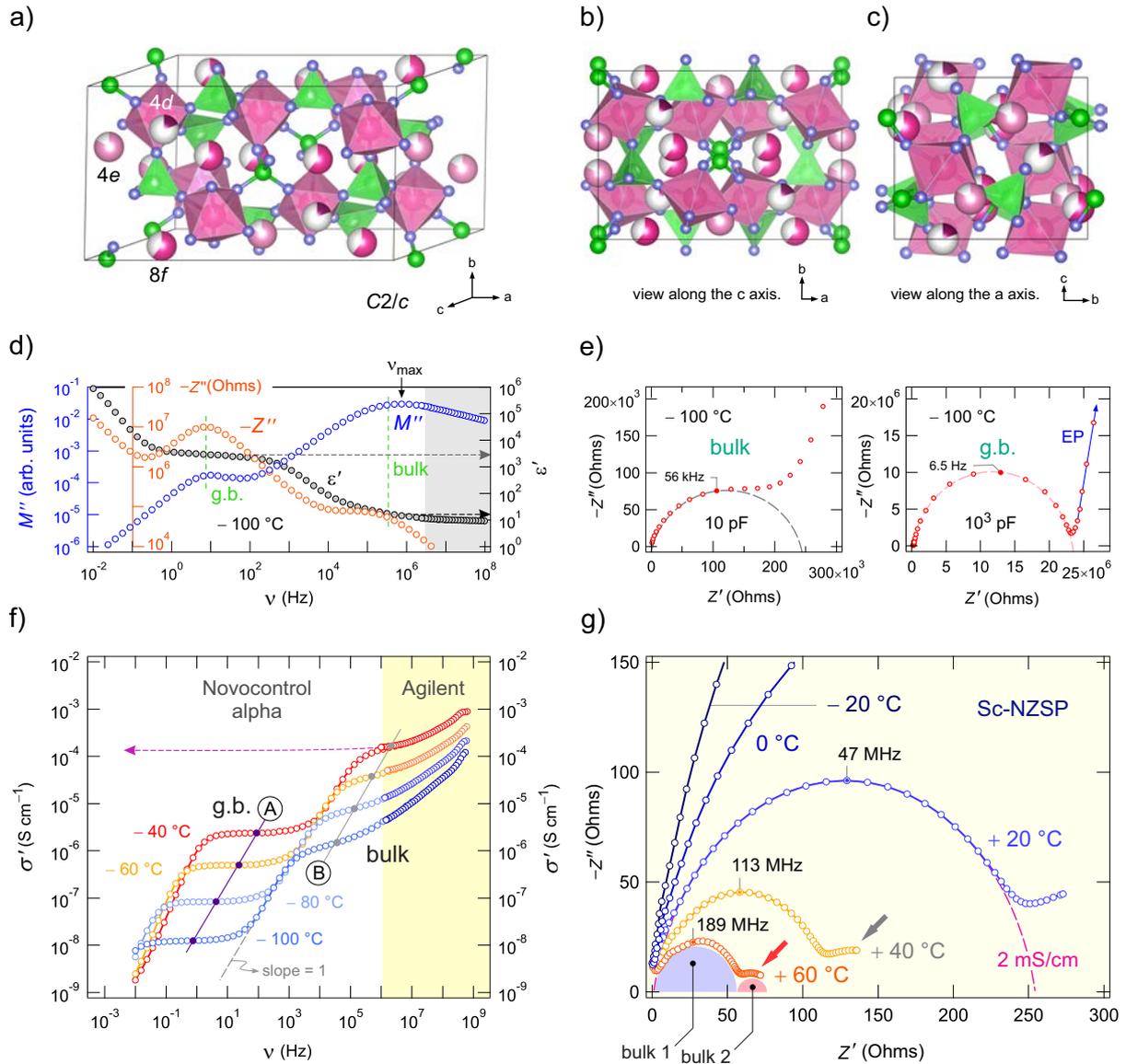

**Figure 1 | Crystal structure and conductivity. a)** to **c)** Crystal structure of monoclinic $Na_3Zr_2(SiO_4)_2PO_4$. **d)** Frequency dependence of the imaginary part of the (i) complex impedance ($-Z''$) and the complex modulus ($M''$) of Sc-NZSP recorded at $-100$ °C. For comparison, the variation of the real part of the complex permittivity is also shown. $-Z''$ and $M''$ each reveal two maxima which correspond to the g.b. ($\varepsilon' = 3 \times 10^3$) and bulk response, respectively; the latter is seen in the high-frequency region, $\varepsilon'$ ranges from 10 to 20. **e)** Complex plane plots of the impedance data measured at $-100$ °C. The depressed semicircle at high frequencies (left) represents the (overall) bulk response ($C_{bulk} = 10$ pF). At intermediate frequencies the grain boundary region appears that is characterized by $C_{g.b.} = 10^3$ pF and a Debye-like non-depressed semicircle with its centre on the $Z'$ axis; piling up of the ions near the blocking electrode, so-called electrode polarization (EP), is seen at the lowest frequencies and manifests as a spike. **f)** Conductivity isotherms of Sc-NZSP recorded at the frequencies and temperatures indicated. We used two impedance analysers to construct the isotherms, a Novocontrol Alpha analyser and an Agilent impedance bridge to acquire the high-frequency data. The frequency independent regions, labelled (A) and (B), represent either g.b. or bulk response. The arrow points to conductivities in the order of $10^{-4}$ S/cm at $-40$ °C. **g)** Nyquist plots recorded at frequencies greater than $10^6$ Hz and at temperatures around ambient. At temperatures above 40 °C the bulk response, which is seen solely here, seems to be composed of at least two semicircles (bulk 1, bulk 2). At room temperature, the intercept of the $-Z''(Z')$ curve with the real axis points

to a conductivity of 4 mS which results, taking the cell constant of the measurement into consideration, in a specific conductivity of 2 mS/cm. See text for further explanation.

Measurements up to the GHz are indispensable to make the bulk response fully visible at temperatures above ambient, see the data points in the GHz range in Figs. **1d** and **1f**. The Nyquist curves in Fig. **1g** show solely impedances measured at frequencies larger than $10^6$ Hz. The curves in the complex plane plots reveal that several semicircles contribute to the overall bulk conductivity; this observation is especially indicated for the Nyquist curves recorded at 40 °C and 60 °C. The substructure of the curves of Sc-NZSP clearly points to multiple, simultaneously occurring electrical relaxation processes with very similar impedances.

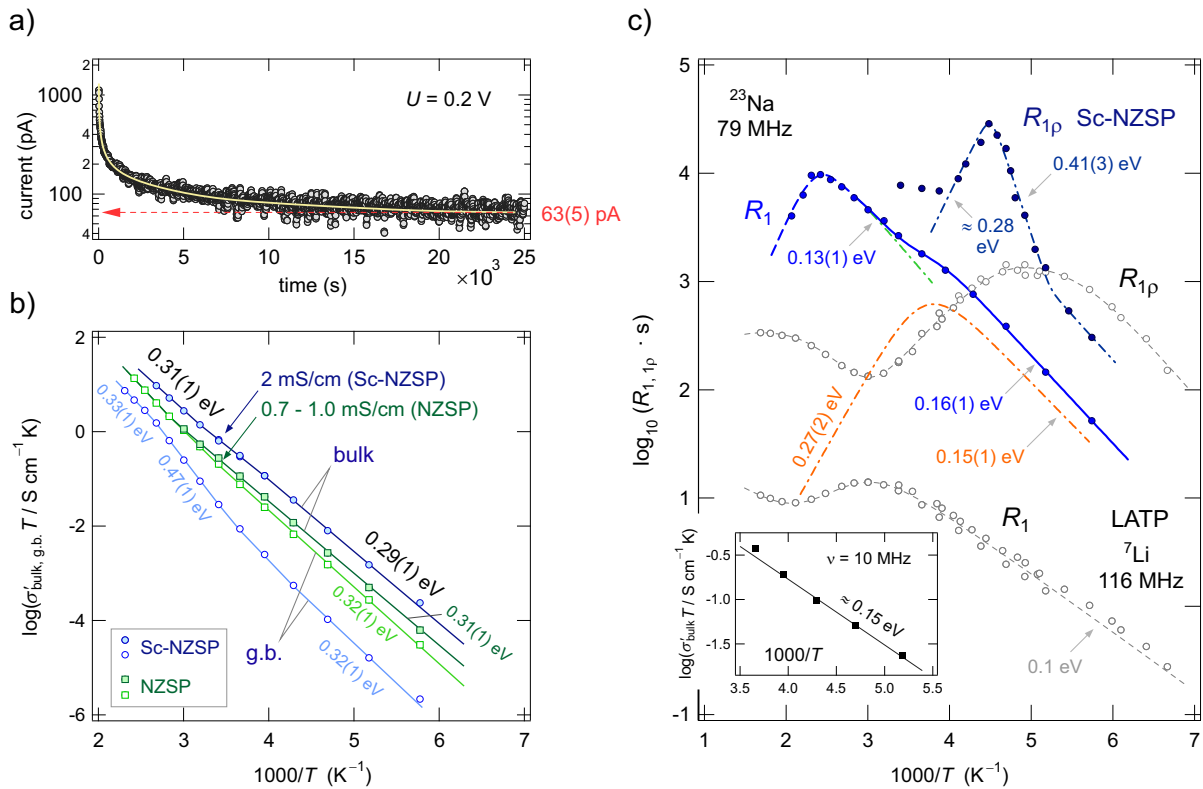

**Figure 2 | Electronic conductivity and temperature dependence of $\sigma'$ of Sc-NZSP, $^{23}$Na NMR spin-lattice relaxation.** **a)** Constant voltage (direct current) polarization curve of a sintered Sc-NZSP pellet with evaporated Au blocking electrodes applied. The high ionic conductivity of the sample translates into a rapid decay of the current at the beginning of the experiment. After a time lapse of sufficient lenght, the current approaches a value that is solely due to the transport of electrons. **b)** Temperature dependence of the bulk and g.b. conductivity of Sc-NZSP ($\sigma_{bulk}$(298 K) = 2.0 mS cm$^{-1}$). The results for NZSP free of Sc are also shown ($\sigma'_{bulk}$(298 K) = 1.0 mS cm$^{-1}$; $\sigma'_{g.b.}$(298 K) = 0.7 mS cm$^{-1}$). Filled symbols refer to bulk values, open ones represent the g.b. response. Activation energies are indicated. Crosses show conductivities taken from Ref. [22]. **c)** $^{23}$Na spin-lattice relaxation rates ($R_1$, $R_{1\rho}$) of Sc-NZSP recorded in both the laboratory frame of reference (79 MHz, $R_1$) and the rotating-frame of reference (20 kHz, $R_{1\rho}$) making use of the spin-lock technique. The sharp diffusion-induced peak of the latter and the shoulder of the $R_1(1/T)$ curve indicate extremely rapid Na ion diffusivity with rates in the order of $\omega_0/2\pi \approx 5 \times 10^8$ s$^{-1}$ at temperatures as low as 250 K. The solid line shows a fit with two BPP-terms to approximate the $R_1$ rates; the dashed-dotted line highlights the asymmetric low-$T$ peak with the activation energies 0.15 eV and 0.27 eV, respectively. The dashed-dotted line drawn through the $R_{1\rho}$ data points is to guide the eye. Unfilled symbols represent, for comparison, results from analogous measurements on a Li-analogue compound, Li$_{1.5}$Al$_{0.5}$Ti$_{1.5}$(PO$_4$)$_3$ (LATP), see ref. [41]. See text for further explanation and discussion.

**Ionic vs. electronic conductivity.** By reading off the conductivities of the plateaus in Fig. **1f**, as indicated by the horizontally drawn arrow, we constructed an Arrhenius plot to analyse the temperature dependence of the two main electrical responses. Potentiostatic polarisation measurements, carried out prior to this analysis, ensured that the overall conductivity analysed is solely determined by ionic charge carriers. For this purpose, we used a symmetric Sc-NZSP pellet outfitted with ion blocking Au electrodes, applied a potential of 0.2 V and followed the current $I$ over time $t$ (Fig. **2a**). After approximately 5 h a stationary state ($I_{t \to \infty}$ = 63(5) pA) is reached in which the concentration gradients are fully developed and the only net charge transport across the material is due to electrons for which the Au contacts are not blocking. The final current of 63 pA (Fig. **2a**) corresponds to an electronic conductivity of $\sigma_{eon, d.c.}$ = 7.2(6) × $10^{-8}$ mS/cm which is lower by 7 to 8 orders of magnitude than the total conductivity $\sigma'$ measured in the limit $\nu \to 0$ for plateau (B), see Fig. **1f**. Thus, Sc-NZSP must be regarded as a pure ionic conductor with the Na ion transference number $t_{ion}$ close to 1.

**Activation energies and solid-state diffusion coefficients.** The overall conductivities plotted in Fig. **2b** reveal that the overall electric transport in Sc-NZSP is determined by activation energies ranging from 0.29 eV to 0.31 eV (–100 °C to 130 °C). The Arrhenius line of the bulk process shows slight deviations from a perfectly linear behaviour. Such a deviation is even more pronounced for the g.b. response for which we identified several regions with activation energies ranging from 0.32 eV to values as high as 0.47 eV. While for temperatures lower than –20 °C bulk and g.b. conductivities differ by two orders of magnitude, above –20 °C $\sigma'_{g.b.}$ increased stronger than expected and was thermally activated with 0.47 eV. At temperatures higher than 120 °C the difference between $\sigma'_{bulk}$ and $\sigma'_{g.b.}$ turned out to be is less than 1 order of magnitude. A ratio $\sigma'_{bulk}/\sigma'_{g.b.}$ close to 1 is, however, observed for NZSP. At 20 °C the bulk conductivity of $Na_3Zr_2(SiO_4)_2PO_4$ is 1.0 mS/cm, for $\sigma'_{g.b.}$ we found 0.7 mS/cm (Fig. **2b**). A conductivity of 2.0 mS/cm for Sc-NZSP results in a room-temperature solid-state diffusion coefficient $D$ of 2 × $10^{-12}$ m²/s if we use the Nernst-Einstein equation to estimate $D$ via $D = \sigma'_{bulk} k_B T/(Nq^2)$ where $k_B$ denotes Boltzmann's constant, $T$ the absolute temperature in K, and $q$ the elementary charge of the Na ions. $N$ refers to the number density of charge carriers. Values in the order of $10^{-12}$ m²/s are comparable with the diffusion coefficients of the best ceramic Li ion conductors known today.

**$^{23}$Na NMR spin-lattice relaxation: Na$^+$ motions on the angstrom scale.** To check whether the high Na$^+$ ion mobility seen in electrical measurements for Sc-NZSP is also reflected in NMR relaxation, we carried out variable-temperature $^{23}$Na spin-lattice relaxation (SLR) measurements in both the laboratory ($R_1$) and rotating frame of reference ($R_{1\rho}$), see Fig. **2c**. The corresponding magnetization transients are shown in Fig. **S1**. While the first are sensitive to diffusive spin fluctuations in the MHz to GHz range, the spin-lock technique allows for the detection of slower motions in the kHz range. As we used a $^{23}$Na NMR Larmor frequency of $\omega_0/(2\pi)$ = 79 MHz to record the $^{23}$NMR SLR rates, we expect the diffusion-induced rate $R_1$ to pass through a maximum at temperatures near or slightly above 20 °C. The jump rate $1/\tau_{NMR}$ at such a maximum is directly determined by the relation $\omega_0 \tau_{NMR} \approx 1$ yielding $D_{NMR} = a^2/(6\tau)$ = 9 × $10^{-12}$ m²/s; for $1/\tau_{NMR}$ we obtain 5 × $10^8$ s$^{-1}$. This estimation uses the Einstein-Smoluchowski equation for uncorrelated 3D diffusion and an average jump distance of 3.3 Å for the various possible exchange processes between the Na sites. In fact, we observe a complex relaxation behaviour with two $R_1$ rate peaks as shown in Fig. **2c**. One of these peaks appears as a shoulder of the main one. It is located at approximately –10 °C if we parameterize the overall relaxation response with a sum of two spectral density functions according to the model of Bloembergen Purcell and Pound [42, 43]. Thus, the low-$T$ peak indicates even faster Na ion exchange processes than seen by conduc-

tivity spectroscopy. The question of whether it is triggered by a possible phase transformation occurring below room temperature and slightly changing the Na$^+$ distribution in Sc-NZSP needs to be checked in further studies.

Conductivity measurements are sensitive to successful Na ion displacements in the limit $\nu \to 0$ of the respective $\sigma'_{bulk}$ plateau (Fig. **1f**); note that the limit $\nu \to 0$ is also often termed the direct current (d.c.) regime, here determined by $E_{a, d.c.}$ = 0.29(1) eV. The very low activation energy $E_{a, NMR}$ of 0.15 eV, see also ref. [44] for comparison) associated with the diffusion-induced NMR rate peak $R_1(1/T)$ seen at lower $T$ (Fig. **2c**) indicate that $^{23}$Na NMR SLR senses the elementary diffusion processes to which local, correlated motions do also contribute. In general, the low-$T$ flank of a given NMR rate peak is influenced by correlation effects and structural disorder also in the form of stress the jumping ions experience while moving in a heterogeneous, irregularly shaped potential landscape [38]. This landscape contains stable as well as transitions states with shorter residence times for the Na$^+$ ions. Forward-backward jumps involving interjacent bottleneck sites [22, 40, 45] result in correlated motion that manifests as dispersive regions in the conductivity isotherms. These dispersive regions are seen in Fig. **1f** in the high frequency region (> 1 MHz) of the bulk isotherms. The activation energy of the high-$T$ flank of the $R_1(1/T)$ peak should be comparable to $E_{a, d.c.}$.

**NMR and conductivity spectroscopy: a comparison.** To compare activation energies from conductivity measurements directly with those from NMR relaxometry we need to analyse alternating current (a.c.) conductivities not in the limit $\nu \to 0$ but in the MHz regime, *i.e.*, on the same time scale where NMR operates. We therefore determined $\sigma'_{bulk}(\nu)$ at $\nu$ = 10 MHz. At this frequency, the dispersive parts are seen fully in the temperature range from –100 °C to –50 °C, which coincides with that of the low-$T$ flank of the first $R_1(1/T)$ peak. A plot of log($\sigma'_{a.c., bulk}$(10 MHz)$T$) vs. 1/$T$ (see the inset of Fig. **2c**) revealed an activation energy $E_{a, a.c.}$ of 0.147(5) eV which is close to that seen by $^{23}$Na NMR SLR in the same temperature range, $E_{a, a.c.}$ = $E_{a, NMR}$. The ratio $E_{a, d.c.}/E_{a, a.c.}$ = 1/$\beta_n$ turns out to be approximately 1.97. On the other hand, $\beta_n$ = 1/1.97 ≈ 0.5 is related to the Jonscher power law exponent [46] describing the dispersive regime of $\sigma'_{bulk}(\nu)$ via the relation $\beta_n$ = 1 – $n$ [47]. If NMR and a.c. conductivity spectroscopy are governed by very similar motional correlation functions in the temperature range given above, $n \approx 0.5$ should be found. In general, $\beta_n$ can be used as a measure to describe the deviation from ideal Debye relaxation that relies on a pure exponential decay function. In the present case, parameterizing the corresponding a.c. isotherms with $\sigma'_{bulk} \propto \nu^n$ indeed yields $n$ = 0.54 indicating similarly shaped correlation functions. This similarity does not contradict the finding that $D_{NMR}$ is larger than $D$ from conductivity measurements. A relatively small correlation factor $f$ ($\leq$ 0.1) linking the two coefficients via $D = f \cdot D_{NMR}$ [48, 49] would explain the difference pointing to highly correlated motions in Sc-NZSP. Most likely, $D_{NMR}$ is indeed influenced by a huge number of unsuccessful, localized jump processes which are not contained in $D$. These localized motions may, however, stimulate long-range ion transport leading to ionic conductivities exceeding that of the Sc-free parent compound with a lower number density of Na$^+$ ions.

An activation energy comparable to $E_{a, d.c.}$ = 0.29 eV is seen in NMR relaxation spectroscopy only if the high-$T$ flank of the diffusion-induced peak $R_1(1/T)$ can be reached [33]. In our case the high-$T$ flank of the spin-lock $R_{1\rho}(1/T)$ peak recorded at a locking frequency $\omega_1/2\pi$ = 20 kHz points to such a value, see Fig. **2c**. However, the peak itself shows anomalous behaviour as the low-$T$ side is characterized by a larger activation energy (0.41 eV) than that on the high-$T$ flank. While such features might be ascribed to low-dimensional diffusion pathways, here the anomaly is supposedly caused by various superimposing jump processes. A point worth mentioning is that the conductivity spectroscopy in the d.c. limit yields an average activation energy of successful net charge transport over a long length scale. The corresponding activation energies extractable from NMR, particularly those in the low-$T$ region of the $R_{1(\rho)}(1/T)$ peaks, specify distinct motional processes with activation energies $E_{a\,low\text{-}T}$ being smaller or even higher than $E_{a\,d.c.}$. The BPP fits shown in Fig. **2c** for the $R_1(1/T)$ behaviour consistently yield a high-$T$ activation energy $E_{a\,high\text{-}T}$ of 0.27 eV being very similar to $E_{a\,d.c.}$, see Table **S3** for the NMR results. The resulting asymmetry of the peak (see supporting information) has frequently been assigned to correlated motion governed by a distribution of (non-)exponential motional correlation functions. The two activation energies of the peak ($E_{a\,low\text{-}T}$ = 0.15 eV and $E_{a\,high\text{-}T}$ = 0.27 eV) are linked to each other via $E_{a\,low\text{-}T} = (\alpha_{NMR} - 1)E_{a\,high\text{-}T}$ yielding $\alpha_{NMR}$ = 1.55; $\alpha_{NMR}$ = 2 corresponding to a quadratic frequency dependence, $R_1 \propto \omega_0^2$, on the low-$T$ side would indicate uncorrelated, isotropic motion. Furthermore, $E_{a\,high\text{-}T}$ (and $E_{a\,d.c.}$) is in line with the activation energy if we analyse the characteristic electric relaxation frequencies $\omega_{max}$ of the Modulus curves in the MHz to GHz regime. $1/\tau_{NMR} = 5 \times 10^8$ s$^{-1}$ agrees perfectly with $\omega_{max}(1/T)$ read off from the $M''$ peaks, see Fig. **S2**. Hence, Modulus spectroscopy and NMR probe the same time scale and sense the same Na$^+$ motional correlation process.

For comparison, in Fig. **2c** also $^7$Li NMR rates of Nasicon-type Li$_{1.5}$Al$_{0.5}$Ti$_{1.5}$(PO$_4$)$_3$ (LATP) are shown [41]. LATP crystallizes with rhombohedral structure and belongs to one the fastest Li-containing phosphates [50]. Although the peak positions of the Na compound studied here are shifted toward somewhat higher temperatures, Sc-NZSP can compete with the high Li$^+$ diffusivity in LATP. As for Sc-NZSP the $^7$Li NMR relaxation response of LATP was analysed with a superposition of at least two diffusion-induced relaxation rate peaks [41].

## Conclusions.

In summary, Na ion dynamics in sol-gel prepared Na$_{3.4}$Sc$_{0.4}$Zr$_{1.6}$(SiO$_4$)$_2$PO$_4$ turned out to be extremely rapid at temperatures around ambient. $^{23}$Na NMR spin-lattice relaxation is governed by complex Na$^+$ dynamics as the relaxation rates reveal several diffusion-induced peaks. While short-range Na motion has to be characterized by activation energies ranging from 0.13 to 0.15 eV, long-range ion transport, on the other hand, follows Arrhenius behaviour with 0.29 to 0.31 eV. Facile Na ion exchange results in room-temperature ion conductivities of 2.0 mS/cm for Na$_{3.4}$Sc$_{0.4}$Zr$_{1.6}$(SiO$_4$)$_2$PO$_4$ and 1.0 mS/cm for the parent compound Na$_3$Zr$_2$(SiO$_4$)$_2$PO$_4$, respectively. Such high bulk conductivities correspond to solid-state diffusion coefficients in the order of $2 \times 10^{-12}$ m$^2$/s. Indeed, for Sc-NZSP $^{23}$Na spin-lattice relaxation NMR points to a self-diffusion coefficient of $9 \times 10^{-12}$ m$^2$/s at 21 °C. Evidently, extremely rapid ion jump processes between regular and intermediate positions contribute to this high value. Such rapid movements trigger the successful jump processes among the different crystallographic sites and facilitate net charge transport over macroscopic distances in Sc-NZSP.

## Methods and experimental design.

**Conductivity analysis.** Sintered pellets with sputtered or evaporated Au electrodes were used to measure complex conductivities at various temperatures and from 0.1 Hz up to the GHz range with a Novocontrol Concept 80 setup. The setup consists of a broadband analyser (Alpha-AN, Novocontrol), which is connected to a BDS 1200 cell in combination with an active ZGS cell interface (Novocontrol) allowing 2-electrode dielectric measurements. The temperature is automatically controlled by means of a QUATRO cryo-system (Novocontrol) making use of a heating element which builds up a specified pressure in a liquid nitrogen Dewar vessel to create a highly constant N$_2$ gas flow. After being heated by a gas jet, the evaporated N$_2$ flows directly through the sample cell that is mounted in a cryostat. With this setup, temperatures can be set with an accuracy of ± 0.01 °C. To reach up to 3 GHz we used an Agilent E4991 A high-frequency analyser connected to a high frequency cell designed by Novocontrol.

**Polarisation measurements**. For the d.c. polarisation measurement a sintered pellet (0.89 mm thick, 7 mm in diameter) was metallised with Au on both sides by evaporation. The pellet was then mounted in an air-tight 2-electrode Swagelok-type cell and connected to the Parstat MC potentiostat equipped with a low-current option. All the preparation steps, including the metallisation, were carried out in Ar or $N_2$ filled gloveboxes filled with an oxygen and water content of less than 1 ppm. The polarisation experiments were performed in a Faraday cage at 23(1) °C.

**$^{23}$Na NMR measurements.** All static $^{23}$Na NMR measurements were carried out using a Bruker Avance III spectrometer equipped with a shimmed cryo-magnet with a nominal magnetic field $B_0$ of approximately 7.04 T. This field strength corresponds to a $^{23}$Na resonance frequency of $\omega_0/2\pi$ = 79.35 MHz. For the NMR measurements, we used sintered pellets which were fire-sealed under vacuum in quartz glass ampoules to permanently protect them from moisture and/or air. We used a Bruker broadband probe (80 mm in diameter) designed for static variable-temperature measurements. The temperature in the sample chamber made of Teflon® was adjusted with a stream of freshly evaporated nitrogen gas. The heating coil was controlled by a Eurotherm unit connected to copper-constantan type T thermocouple placed in the immediate vicinity of the sample. To study Na ion dynamics, $^{23}$Na NMR spin-lattice relaxation experiments in both in the laboratory and in the rotating frame were carried out with pulse sequences using suitable phase cycling to suppress unwanted coherences. $R_1$ rates were recorded with the saturation recovery pulse sequence consisting of a train of 90° pulses to destroy any longitudinal magnetization $M$. The subsequent recovery of $M$ was recorded after variable delay times $t_d$ with a 90° reading pulse. The $\pi/2$ pulse lengths ranged from 2 to 4 µs at 200 W, depending on temperature. Up to 64 scans were accumulated to obtain $M(t_d)$ for each waiting time. The area under the free induction decays was used as a measure for $M(t_d)$. The magnetization transients $M(t_d)$ were analysed with stretched exponential functions to extract the rates $R_1$; stretching exponents varied from 0.8 to 1. The spin-lock measurements were recorded with a locking frequency of 20 kHz. At the beginning of the pulse sequence a 90° pulse prepares the spin system. Immediately after this, the locking pulse with variable duration $t_{lock}$ (46 µs to 46 ms) was used to observe transversal relaxation of the flipped magnetization $M_\rho$. The decay of $M_\rho$ in the $(xy)_\rho$-plane of the rotating frame of reference is detected with a 90° reading pulse. The number of scans for each data point of the magnetization curve was 64. A recycle delay of at least $5 \times 1/R_1$ ensured full longitudinal relaxation. $M_\rho(t_{lock})$ could be fitted with stretched exponentials, stretching factors ranged from 0.8 to 0.4 depending on temperature.

**Acknowledgements.** We thank our colleagues at the TU Graz for valuable discussions. Financial support from the Deutsche Forschungsgemeinschaft (DFG Research Unit 1277, grant no. WI3600/2-2 and 4-1) as also from the Austrian Federal Ministry of Science, Research and Economy, and the Austrian National Foundation for Research, Technology and Development (CD-Laboratory of Lithium Batteries: Ageing Effects, Technology and New Materials) is greatly appreciated.

**Contributions.** S.L. and D.P. contributed equally to this work. All authors conceived and supervised the project and also contributed to the interpretation of the data.

**Competing interests.** The authors declare no competing financial interests.

**Corresponding author.** Correspondence to M. Wilkening.